\RequirePackage{fix-cm}

\documentclass[twocolumn]{svjour3}          
\smartqed  
\usepackage{graphicx}
\usepackage[misc]{ifsym}
\usepackage{amsmath}
\usepackage{amssymb}
\usepackage{subfigure}

\begin{document}

\title{The Impact of Vaccination Behavior on Disease Spreading Based on Complex Networks}

\author{Yingyue Ke\and
        Jin Zhou 
}

\institute{Yingyue Ke\and J. Zhou $(\textrm{\Letter})$ \and authour3 \at
	School of Mathematics and Statistics, Wuhan University, Wuhan 430072, China\\
	\email{jzhou@whu.edu.cn}\\
	~~~\\
	J. Zhou\\
	Hubei Key Laboratory of Computational Science, Wuhan University, Hubei 430072, China          
}

\maketitle

\begin{abstract}
	
     Vaccination is an effective way to prevent and control the occurrence and epidemic of infectious diseases. However, many factors influence whether the residents decide to get vaccinated or not, such as the efficacy and side effects while individuals hope to obtain immunity through vaccination. In this paper, the public attitude toward vaccination is investigated, especially how it is influenced by the public estimation of vaccines efficacy and reliance on their neighbors' vaccination behavior. We find that improving people's trust in the vaccination greatly benefits increasing the vaccination rate and accelerating the vaccination process. Counterintuitively, if the individual's attitude towards vaccination is more reliant on his neighbors' vaccination behavior, more individuals will get vaccinated, and the vaccination process will speed up. Besides, individuals are more willing to get vaccinated if they have more neighbors.
	
	\keywords{ disease spreading \and vaccination \and complex network  }
	
\end{abstract}

\section{Introduction}

Since the coronavirus disease 2019(COVID-19) broke out in Wuhan at the end of 2019, it has deprived countless lives and done great harm to thousands of families in the past two years. COVID-19 has imposed tremendous pressure on the healthcare system in all regions and countries. Without any exaggeration, some public medical systems have almost been paralyzed during the severe epidemic. This ongoing epidemic is destined to be a heart-wrenching disaster \cite{Nishi2020,Hui2020,Tang2020}.

Epidemic dynamics has always been a topic of great concern to scholars. Researchers have always been managing to establish various models to explore the internal mechanism transmission of different kinds of diseases, predict the spreading of the disease and design practical strategies to control epidemics \cite{Liu2019,Gross2006,Newman2002}. Complex Networks is a crucial theoretical tool to study various large-scale and complex systems in the real world \cite{Zhu2020}. Before the outbreak of COVID-19, scholars have accumulated much experience on the spreading of infectious diseases. They have researched transmission mechanisms of epidemics with different models based on various networks \cite{Wei2018,Sun2018,Wang2019,Granell2013}. After the COVID-19 broke out as an unknown, highly contagious, and highly harmful disease, it quickly attracted the attention of scholars all over the world \cite{Bert2020,Ivorra2020}. Researchers try to construct proper models to understand its transmission mechanism and predict its future development to propose favorable methods to control the spread and reduce the loss based on past experiences and existing data \cite{NiuR2020,Zhan2020}. At the same time, scholars propose various methods to reduce contact between people and try to suppress the increase in the number of infected individuals because COVID-19 can be transmitted from person to person \cite{Liw2020,Pastor2003}.

When the vaccine against COVID-19 came out, the governments encouraged residents to get vaccinated to protect residents from COVID-19. Vaccination has an absolute advantage in preventing individuals from being infected \cite{Zhou2011,Lv2020,Lwq2021}. However, the vaccine is new, which means it has not undergone a long-term and stable clinical test. On the one hand, people are well aware of the advantage of vaccination. On the other hand, they always have doubts on the efficacy and side effects of this new vaccine. Therefore, although the vaccine against COVID-19 is qualified and the government motivates people to get vaccinated, not everyone is willing to do it. Based on these analyses, In this paper, we mainly study how an awareness that the individual hopes to gain immunity from vaccination but does not want to bear the side effects vaccination may cause influences individuals' vaccination behavior. A reliable prerequisite to realizing this scenario is that all members around the individual should be vaccinated. More specifically, if all neighbors are vaccinated, then an individual will be in a safe encirclement, and thus there is no need to get vaccinated for the individual himself. This awareness can explain why many people pay attention to their neighbors's attitude towards the vaccination with the thought that if all my neighbors get vaccinated, then I do not need to do it.

In the paper, we attempt to investigate two questions. The first one is how the trust in vaccines affects the vaccination process ? The second is how does reliance on neighbors influences individuals' willingness to get vaccinated? The greater reliance on neighbors means that every time the neighbor's vaccination behavior changes, the vaccination willingness for the individual himself will fluctuate more significantly. Results show that increasing trust in vaccines or reliance on the vaccination behavior of neighbors will increase the vaccination rate and hasten the vaccination process. And increasing trust in vaccines will cause a more significant improvement on the vaccination progress. Besides, individuals with more neighbors tend to have a stronger willingness to get vaccinated.

The remaining part of the paper is organized as follows. In Sec. \uppercase\expandafter{2}, we describe the process of vaccination reliant on neighbors. The main conclusion is in Sec. \uppercase\expandafter{3}. Finally, the main work in the paper is summarized, and its practical significance is discussed.

\begin{figure*}
	\centering
	\includegraphics[width=\textwidth]{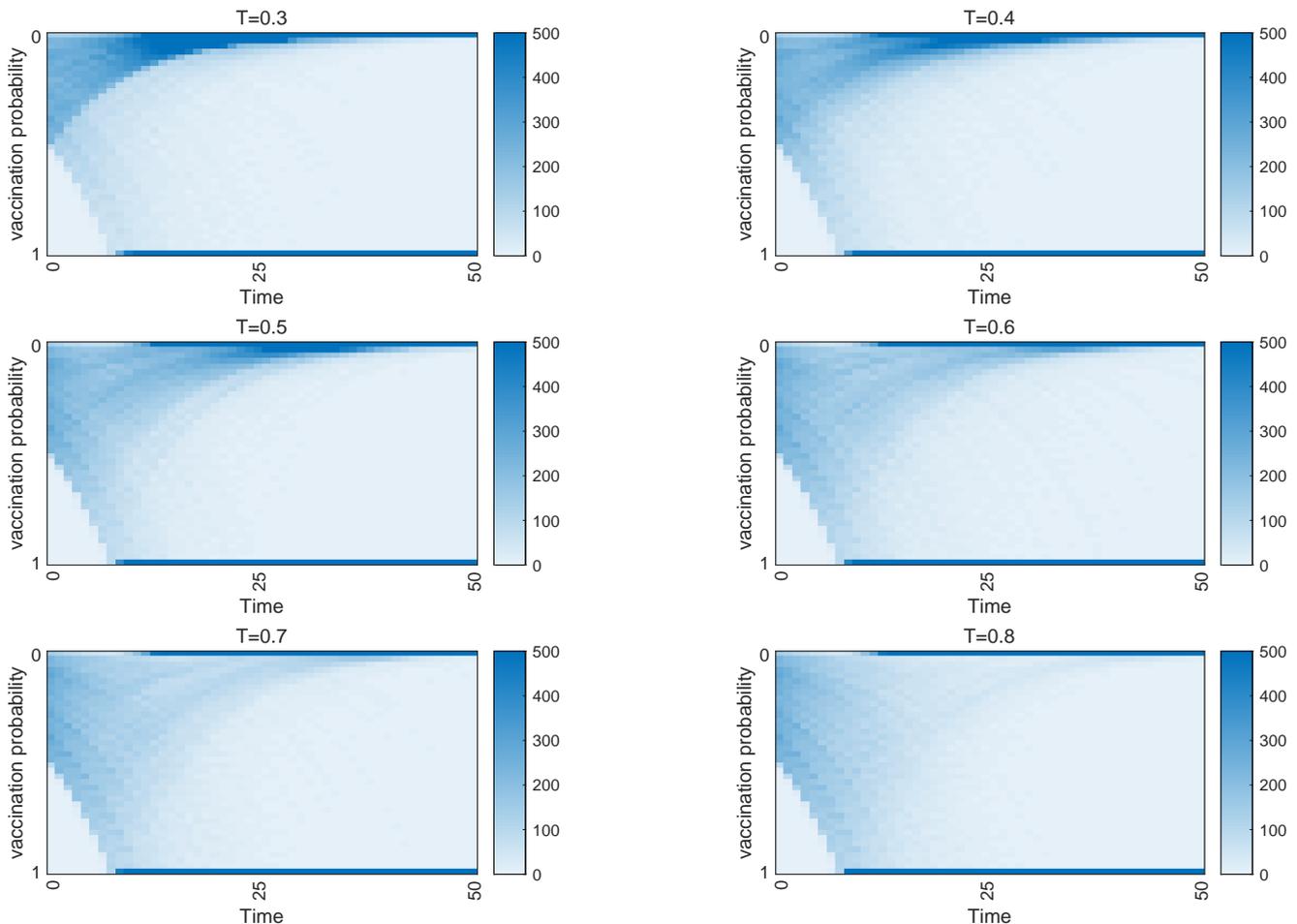}
	\caption{The time evolution of the proportion of nodes in different vaccination willingness under several different values of $ T $ with $ r=0.1 $. $ T=0.3, 0.4, 0.5, 0.6, 0.7$ and $ 0.8 $.}\label{fig1}
\end{figure*}

\begin{figure*}
	\centering
	\includegraphics[width=\textwidth]{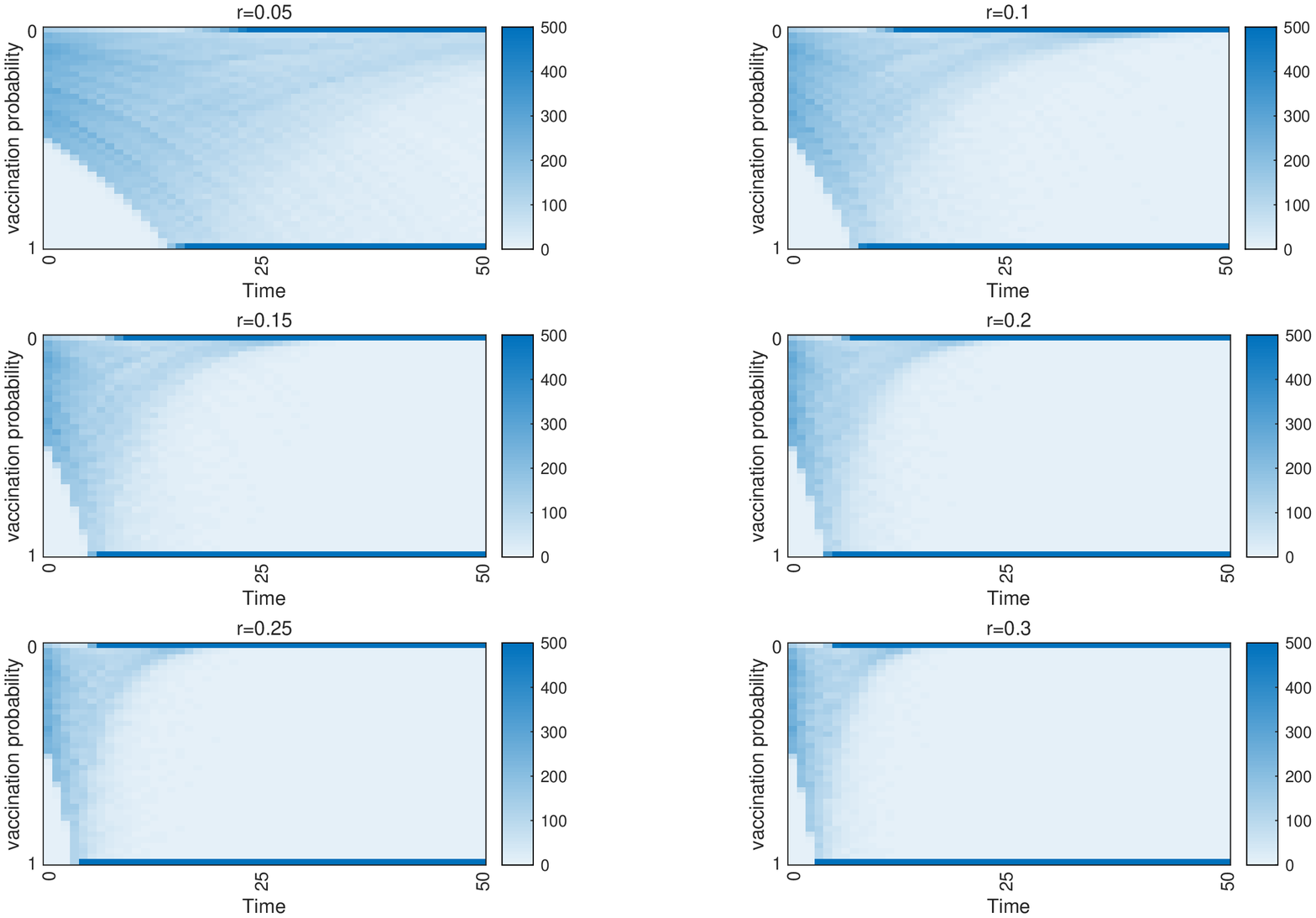}
	\caption{The time evolution of the proportion of nodes in different vaccination willingness under several different values of $ r $ with $ T=0.7 $. $ r=0.05, 0.1, 0.15, 0.2, 0.25$ and $ 0.3 $.}\label{fig2}
\end{figure*}

\section{Model}
\label{sec:2}

In this paper, we construct a model to describe how the vaccination process evolutes with considering the efficacy of vaccines and the influence of neighbors' vaccination behaviors. The model is based on a network. The relationship in a community could be described by an adjacency matrix $ A_{N \times N} $, where $ N $ denotes the size of the network and means the number of individuals in the community, and each node represents an individual. If $A_{ij}=1$, then node $ i $ is connected to node $ j $, which means that the individual $ i $ and the individual $ j $ are neighbors. Their attitude towards vaccination has a direct influence on each other; otherwise, $A_{ij}=0 $, then node $ i $ is not connected to node $ j $, which means that the individual $ i $ and the individual $ j $ are not neighbors. Their willingness to get vaccinated indirectly impacts each other at most. $ N_{i}=\sum_{j=1}^{N}A_{ij} $ represents the number of neighbors for individual $ i $.

The vaccination evolution model includes two main factors, the trust in vaccines and the reliance on neighbors. Take individual $ i $ as an example. The trust in vaccines $ T_{i} \in [0,1] $ reflects the estimation of the efficacy of the vaccine, with higher $ T  $ indicating that the individual $ i $ believes the vaccine is more effective and has fewer side effects. The reliance on neighbors $ r_{i} $ reflects the extent to which neighbours' willingness to get vaccinated causes fluctuations in the individual $ i $ willingness of vaccination, with higher $ r_{i}  $ meaning greater fluctuations in the individual $ i $ willingness to get vaccinated. The willingness of vaccination is denoted by $ P_{i}(t_{n}) $ $ (  i=1,2,\ldots,N, n=0,1,2,\dots ) $, which represents the willingness to get vaccinated for individual $ i $ at time $ t_{n} $. 
The initial value is
\begin{equation}\label{f1}
P_{i}(t_{0})=\frac{N_{i}}{N_{max}}+\delta_{i},
\end{equation}
where $ N_{max} $ represents the most neighbors that an individual has in the network, and $ \delta_{i} \in [0,1] $ is the disturbance, representing other factors that influence the individual's willingness to get vaccinated at beginning. From equation (\ref{f1}), we see that when the vaccine comes out, the more neighbors an individual has, the more likely the individual is to get vaccinated. The reason is that more neighbors mean a greater risk of being infected. Thus, people will have a stronger willingness to get vaccinated to avoid the risk of being infected when they have more neighbors.

At each time $ t_{n} $, every individual could choose to increase the willingness to get vaccinated or decrease the willingness to get vaccinated or keep the same as last time. That is,

\begin{equation}
P_{i}(t_{n})=P_{i}(t_{0}) \prod_{j=1}^{n}{(1+r_{i}\cdot sgn(T_{i}-AP_{i}(t_{j-1})))}
\end{equation}
$ n=1,2,3,\dots $,where $ r_{i} $ represents the reliance on neighbors of the individual $ i $, $ T_{i} $ represents the trust in vaccines for the individual $ i $, and
\begin{equation}
AP_{i}(t_{j-1})=\left( \sum_{m=1}^{N_{i}}{P_{i}^{m}(t_{j-1})} \right) /N_{i}
\end{equation}
represents the average willingness of individual $ i' $ neighbors to get vaccinated. Here, it is worth noting that $ P_{i}(t_{n}) \in [0,1] $ $ (  i=1,2,\ldots,N, n=0,1,2,\dots ) $, if $ P_{i}(t_{n})\geqslant 1 $ at any time $ t_{n} $, then $ P_{i}(t_{k})\triangleq 1$, $\forall\ k \geqslant n  $. This means that once $  P_{i}(t) $ reaches 1, then the individual $ i $ will go to get vaccinated. So the individual has been vaccinated and its status will not change from then on.

\begin{figure}
	\centering
	\subfigure[]{\label{3a}\includegraphics[width=0.45\textwidth]{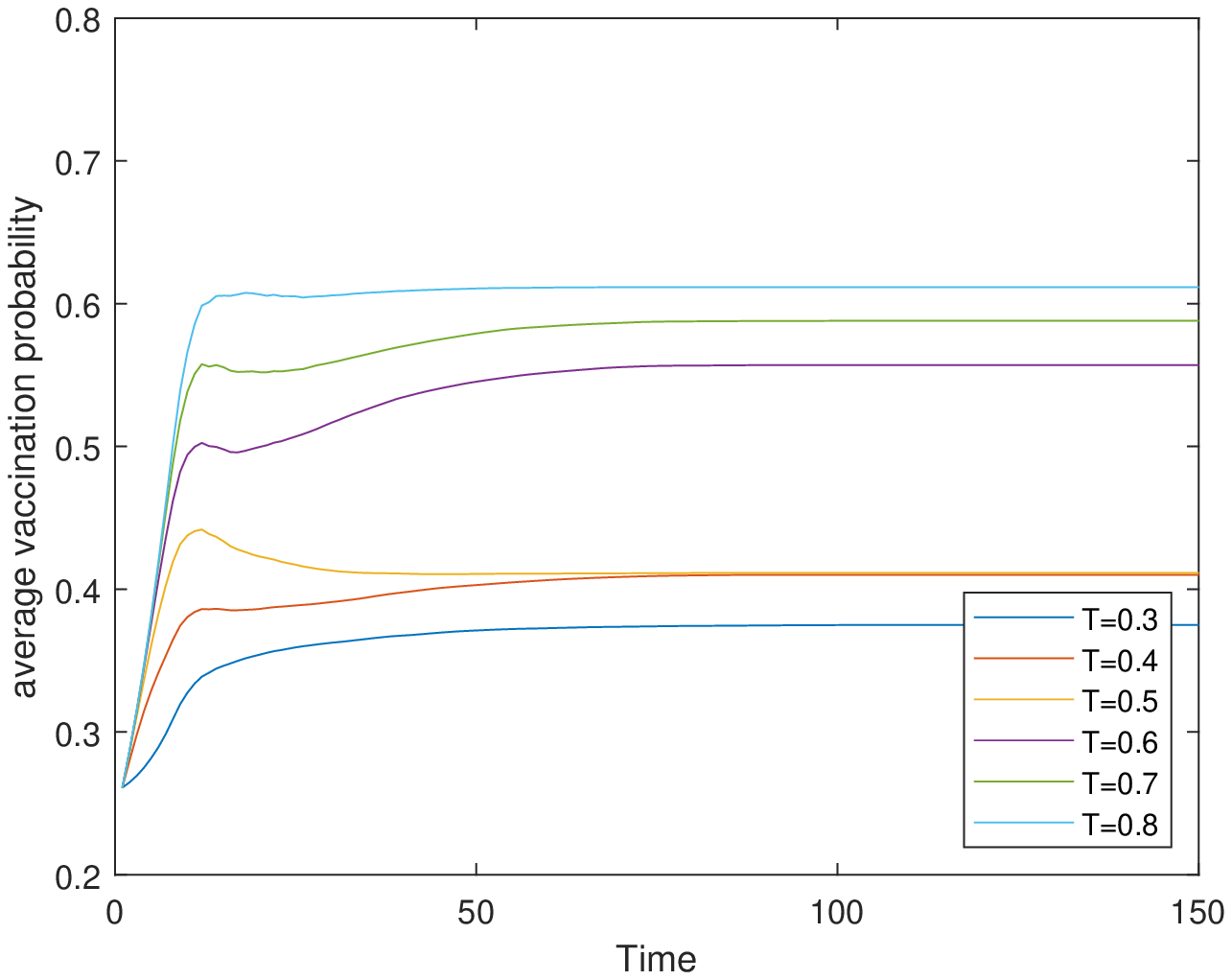}}
	\subfigure[]{\label{3b}\includegraphics[width=0.45\textwidth]{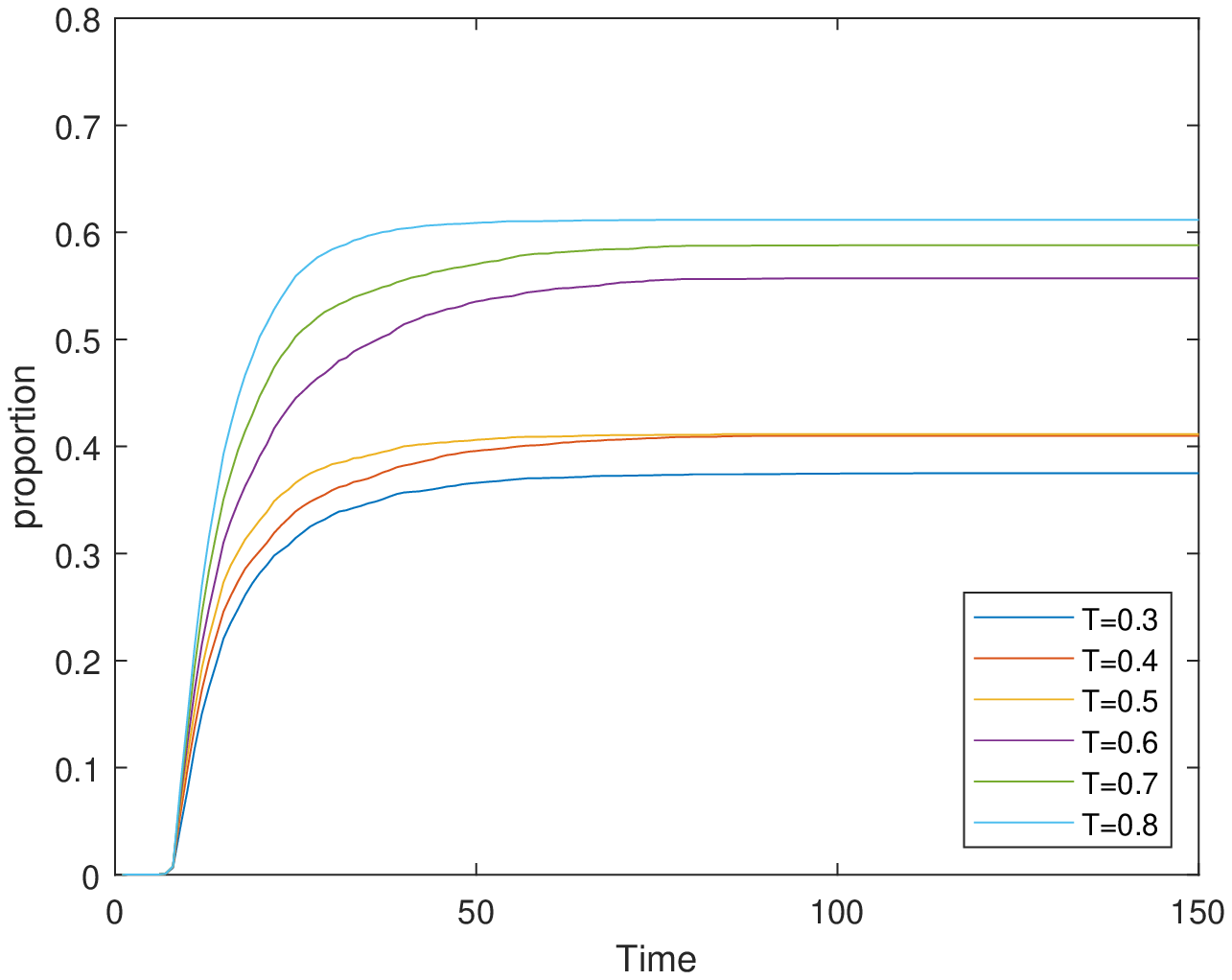}}
	\caption{(a) The time evolution of average vaccination willingness of all nodes under several different values of $ T $ with $ r=0.1 $. Lines with different color represents different value of $ T $ as shown. $ T=0.3, 0.4, 0.5, 0.6, 0.7$ and $ 0.8 $. (b) The time evolution of vaccinated nodes proportion under several different values of $ T $ with $ r=0.1 $. Lines with different color represents different value of $ T $ as shown. $ T=0.3, 0.4, 0.5, 0.6, 0.7$ and $ 0.8 $.}\label{fig3}
\end{figure}

\begin{figure}
	\centering
	\subfigure[]{\label{4a}\includegraphics[width=0.45\textwidth]{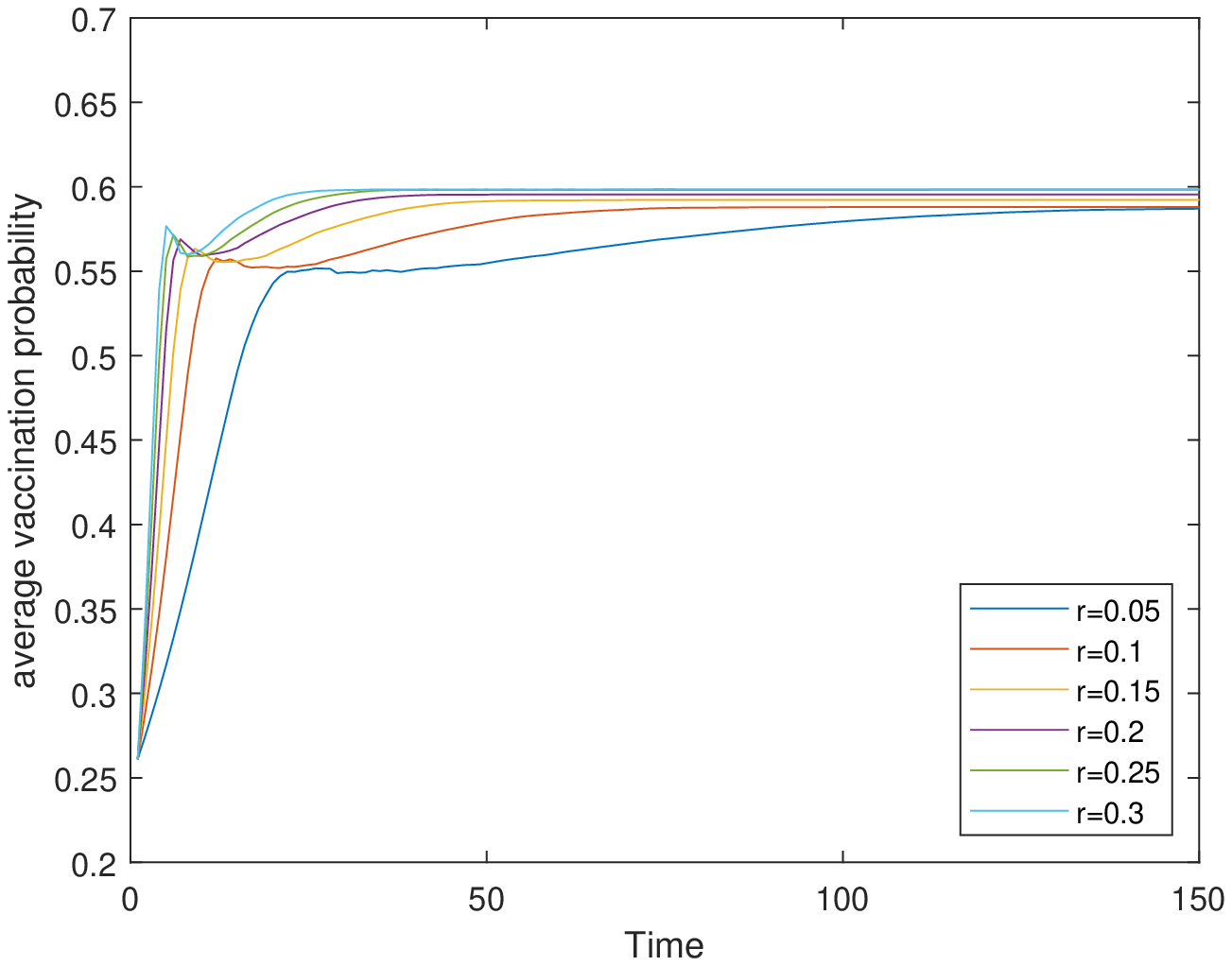}}
	\subfigure[]{\label{4b}\includegraphics[width=0.45\textwidth]{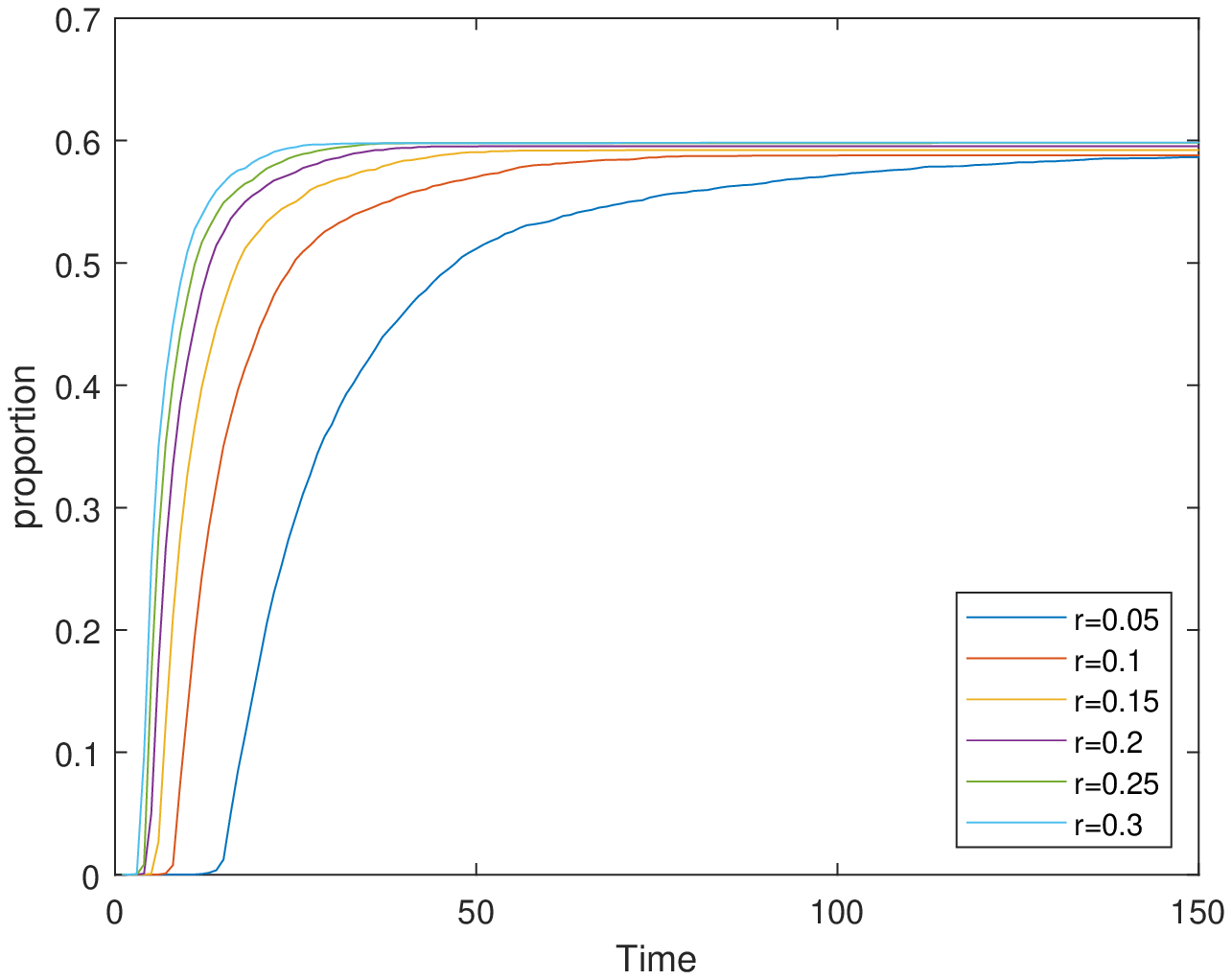}}
	\caption{(a) The time evolution of average vaccination willingness of all nodes under several different values of $ r $ with $ T=0.7 $. Lines with different color represents different value of $ r $ as shown. $ r=0.05, 0.1, 0.15, 0.2, 0.25$ and $ 0.3 $. (b) The time evolution of vaccinated nodes proportion under several different values of $ r $ with $ T=0.7 $. Lines with different color represents different value of $ r $ as shown. $ r=0.05, 0.1, 0.15, 0.2, 0.25$ and $ 0.3 $.}\label{fig4}
\end{figure}

\section{Results}
\label{sec:3}
In this section, we perform a mass of stochastic simulations to show the impact of neighbor-reliant awareness and the efficacy of vaccines on the vaccination. Before that, we make some necessary statements about the network. We build a BA network to approximate the relationship between individuals in a community. Initially, we introduce $ m_{0}=200 $ nodes. The new nodes continue to be linked to the existing network with each $ m=5 $ until the network size reaches $ N=5000 $. Then all the simulations are based on this network. In order to simplify the simulation, we let all $ T_{i} $ and $ r_{i} $ take the same value as $ T $ and $ r $ in the simulation.

First, we focus on the impact of trust in vaccines on vaccination. It is seen from Fig. \ref{fig1} that, as the trust in vaccines $ T $ increases, nodes clustering in low vaccination willingness become less and gradually approach to higher vaccination willingness, and the distribution of nodes tends to stabilize faster. That is to say, if individuals trust vaccines more, individuals' willingness to get vaccinated will be stronger, and the entire vaccination process will speed up.
In addition, how the average vaccination willingness of the whole network $ AP(t_{n})=\sum_{i=1}^{N}P_{i}(t) /N $ changes over time presents three different trends as it is shown in Fig. \ref{3a}. When the trust in vaccines $ T $ is relatively small, the average vaccination willingness to get vaccinated of all individuals will rapidly rise to a peak at first and then slowly rise until it stabilizes at a relatively small level. When the trust in vaccines $ T $ is moderate, the average willingness will rapidly increase to a higher peak and gradually decrease until it stabilizes at a relatively higher level. When the trust in vaccines $ T $ is enough great, the average vaccination willingness will first rapidly increase to a much higher peak, then go through a short period of decline, and then slowly increase until it stabilizes at a much higher level.
Also, from Fig. \ref{3b}, it is clear that the final fraction of vaccinated individuals $ VP(t_{n})=\sum_{i=1}^{N}P_{i}(t_{n}) /N, P_{i}(t_{n})=1 $ is increasing and stabilizes more rapidly as the trust in vaccines $ T $ is increases. And its evolution keeps similar under different $ T $, which rises fast at first and then gradually stabilizes.
It is worth noting that even a period of decline in the middle will occur when the trust in vaccines $ T $ is great. However, the initial peak of the average vaccination willingness and the final vaccination rate will increase with the increase of the trust in vaccines $ T $. The vaccination process will also speed up under a higher the trust in vaccines $ T $. It shows that enhancing the trust in vaccines is an effective way to promote the rate of vaccination even though people will experience a short time of willingness decreasing during the process.

\begin{figure}
	\centering
	\subfigure[]{\label{5a}\includegraphics[width=0.45\textwidth]{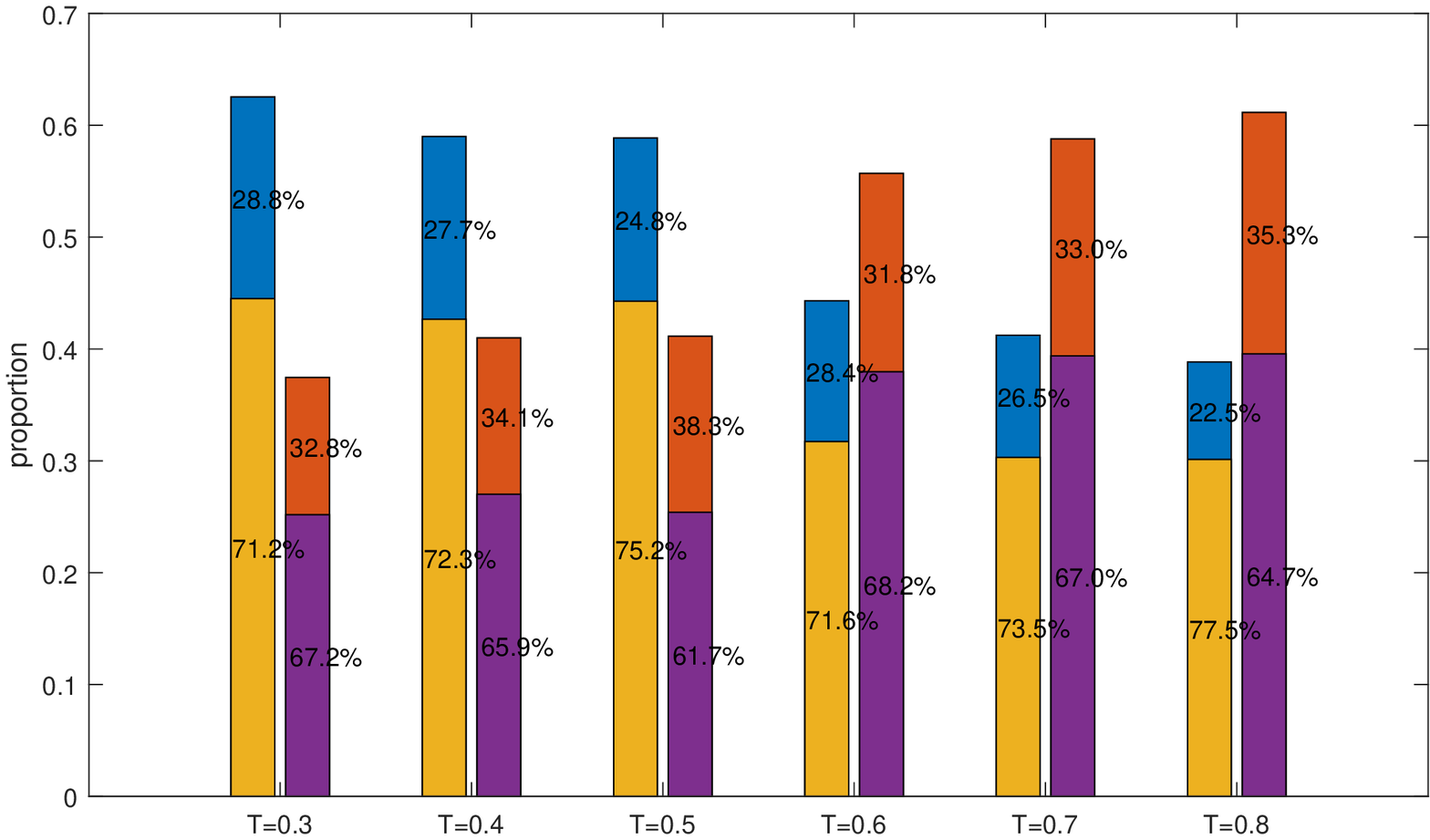}}
	\subfigure[]{\label{5b}\includegraphics[width=0.45\textwidth]{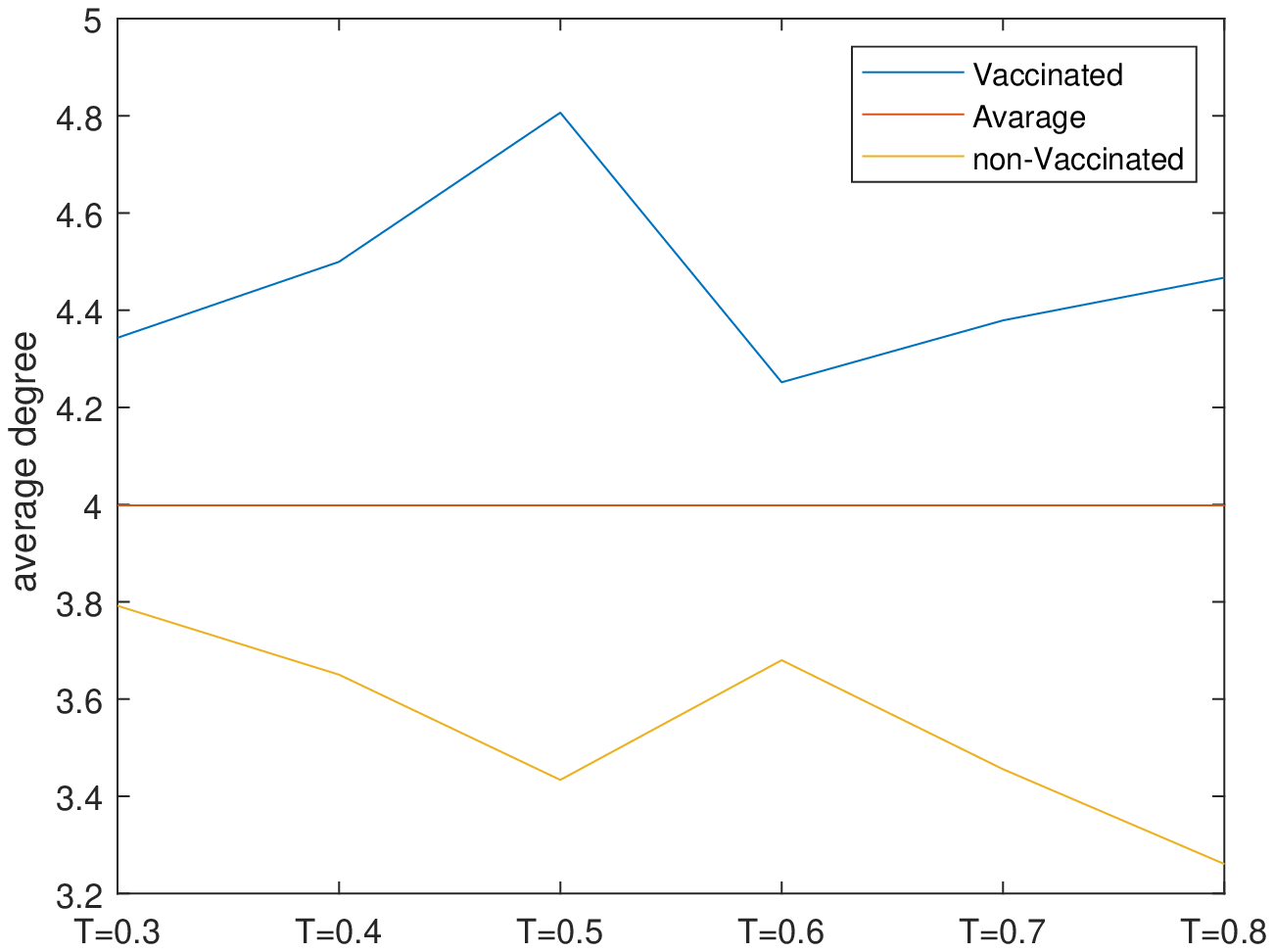}}
	\caption{(a) The final proportion of nodes whose degrees are higher than the average degree or lower than the average degree in the vaccinated nodes and the non-vaccinated nodes under $ T=0.3, 0.4, 0.5, 0.6, 0.7$ and $ 0.8 $ with $ r=0.1 $. The blue, yellow, orange, and purple accounts for the fraction of nodes: non-vaccinated and the degree is higher than the average degree, non-vaccinated and the degree is lower than the average degree, vaccinated and the degree is lower than the average degree, vaccinated and the degree is higher than the average degree, respectively. (b) The final average degree of vaccinated nodes and non-vaccinated nodes under $ T=0.3, 0.4, 0.5, 0.6, 0.7$ and $ 0.8 $ with $ r=0.1 $.  }\label{fig5}
\end{figure}

\begin{figure}
	\centering
	\subfigure[]{\label{6a}\includegraphics[width=0.45\textwidth]{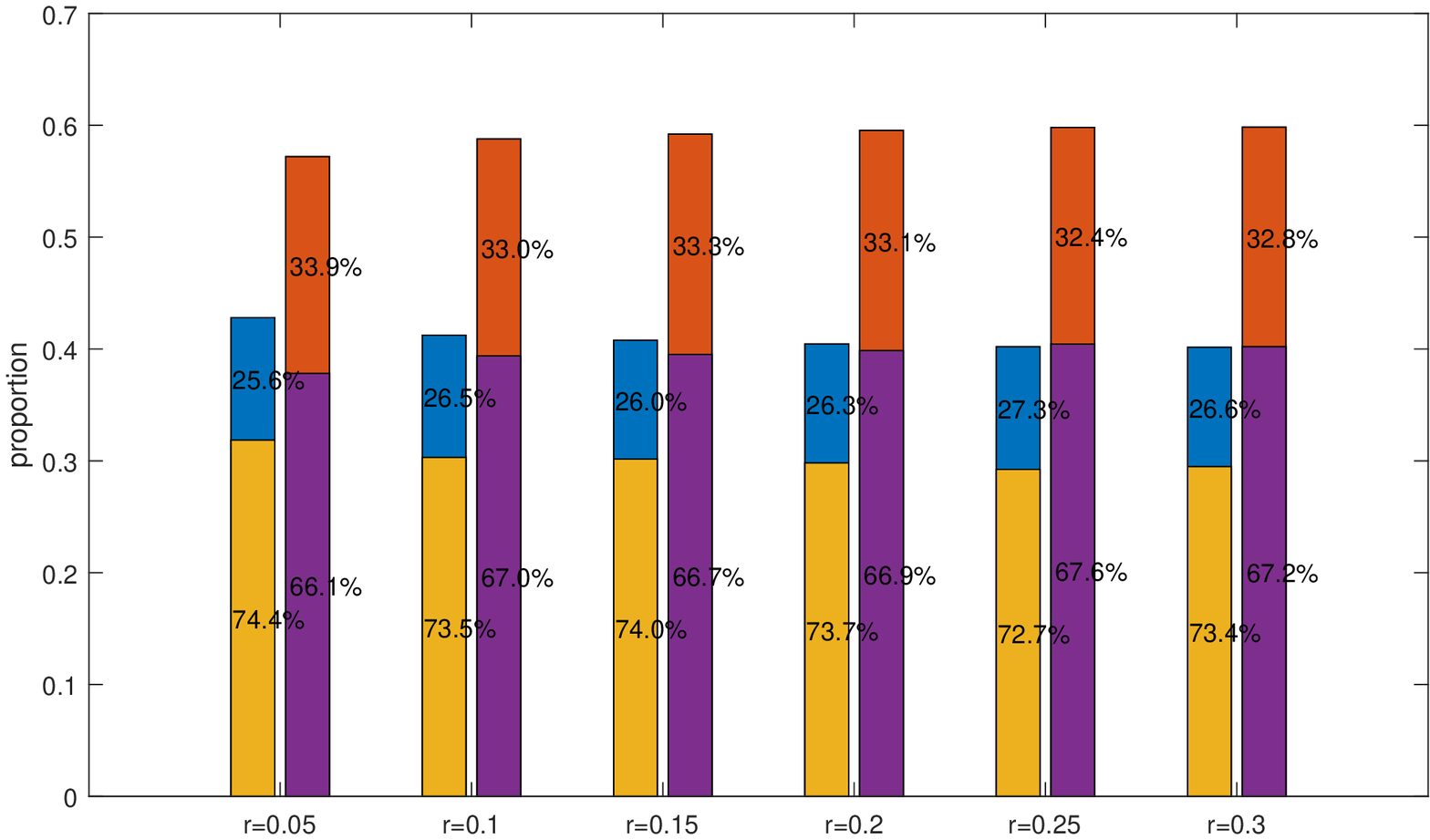}}
	\subfigure[]{\label{6b}\includegraphics[width=0.45\textwidth]{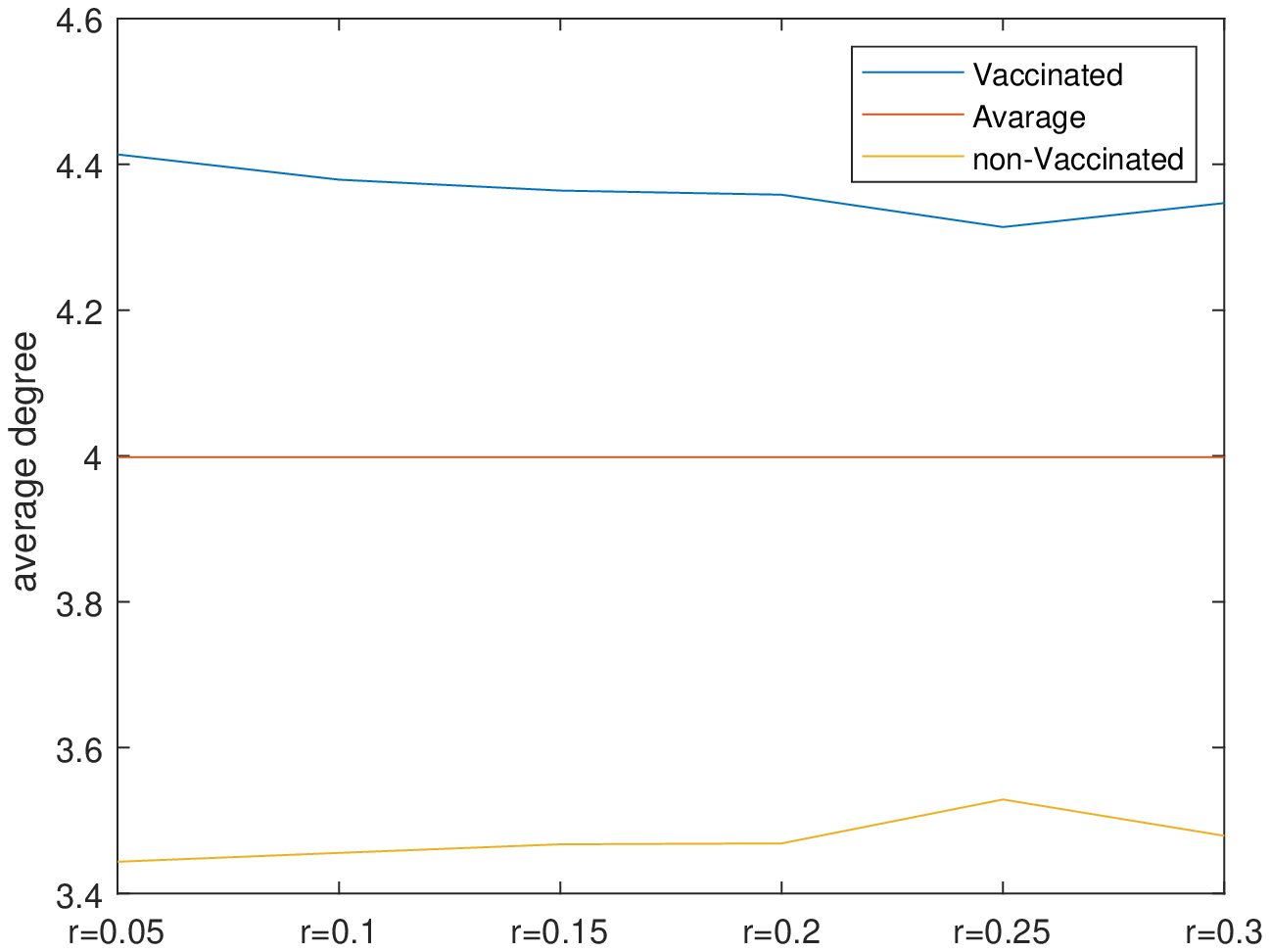}}
	\caption{(a) The final proportion of nodes whose degrees are higher than the average degree or lower than the average degree in the vaccinated nodes and the non-vaccinated nodes under $ r=0.05, 0.1, 0.15, 0.2, 0.25$ and $ 0.3$ with $T=0.7 $. The blue, yellow, orange and purple accounts for the fraction of nodes, who are non-vaccinated and the degree is higher than the average degree, non-vaccinated and the degree is lower than the average degree, vaccinated and the degree is lower than the average degree, and vaccinated and the degree is higher than the average degree, respectively. (b) The final average degree of vaccinated nodes and non-vaccinated nodes under $ r=0.05, 0.1, 0.15, 0.2, 0.25$ and $ 0.3$ with $T=0.7 $. }\label{fig6}
\end{figure}

To clarify how reliance on neighbors affects the evolution of vaccination, we explore similar characters under different reliance on neighbors $ r $ with the trust in vaccines $ T $ fixed. From Fig. \ref{fig2}, we know that as the reliance on neighbors $ r $ increases, less individuals cluster in the area where the the willingness to get vaccinated is small, and approache to the area where the vaccination willingness is higher and the vaccination process stabilizes much faster. 
Different from the influence of $ T $, the increase of the reliance on neighbors $ s $ does not change the trend of the time evolution of average vaccination willingness $ AP $. From Fig. \ref{4a}, whatever the reliance on neighbors $ s $ is, $ AP(t_{n}) $ will first rapidly increase to a peak, then go through a short period of decline, and then slowly increase until it stabilizes to a higher level. Also, a greater reliance on neighbors $ s $ deduces a higher peak and a higher final level of average vaccination willingness $ VP(t_{n}) $. Besides, \ref{4b} tell us that if individuals are more reliant on neighbors, there will be more individuals willing to get vaccinated, and the process of vaccination will speed up.

Finally, we investigate the influence of degree on the vaccination. In the first step, we analyze the final proportion of individuals with a degree greater than the average degree or lower than the average degree in the vaccinated ones and in the non-vaccinated ones under different $ T $ and $ r $ respectively. We observe from Fig. \ref{5a} and Fig. \ref{6a} that compared in non-vaccinated individuals, the proportion of individuals with a degree greater than the average degree is higher in vaccinated ones. In the second step, we analyze The average degree of vaccinated nodes and non-vaccinated nodes under different $ T $ and $ r $. It is seen from Fig. \ref{5b} and Fig. \ref{6b}  that the average degree of non-vaccinated nodes is significantly lower than the average degree, but the average degree of vaccinated nodes is obviously higher than the average degree of. The above results mean that the more neighbors people have, the more they are more willing to get vaccinated because the more people they come into contact with every day, the greater the chance of being infected by their neighbors is. Also, more neighbors mean it is harder to realize neighbor-reliant immunity(NRI) because it means more neighbors need to get vaccinated if the individual wants to be in a safe encirclement. Thus, people have a stronger willingness to get vaccinated with more neighbors.

\section{Conclusion}
In this article, we have simulated the vaccination process under the influence of the combination of the trust in vaccines and the reliance on neighbors. When the neighbor's average vaccination willingness is lower than an individual's trust in the vaccine, the individual will be more likely to consider the immunity with vaccination and therefore more willing to get vaccinated. Otherwise, the individual is more likely to think about the side effects of vaccination and IS less likely to get vaccinated. Numerical simulations show that when people increase their trust in vaccines or their reliance on neighbors, the final vaccination rate will increase and the vaccination process will speed up, but increasing the trust in vaccines faciliates the vaccinatin process more. Besides, individuals with more neighbors tend to have a stronger willingness to get vaccinated, so the percentage of individuals with a degree higher than the average degree of the network in vaccinated ones is higher than in non-vaccinated ones.
The significance of these conclusions is that we could improve the vaccination rate and vaccination speed by adopting some methods to improve people's trust in vaccines and reliance on neighbors. And, it is more effective to increase people's trust in vaccines than to increase people's reliance on neighbors.

In this paper, an individual's decision of vaccination is mainly determined by the average vaccination willingness of its neighbors. In fact, the process is more complicated, and the change of vaccination willingness will be affected by other important factors, for example, the number of vaccines. If the number of vaccines is limited which not everyone is accessible to get vaccinated when they want, it may increase the willingness to get vaccinated. For another example, the government may require residents to get vaccinated. If not, their travel may be restricted, and so on. Besides, the research based on a multiplex network is worthy of in-depth consideration.

\begin{acknowledgements}
	This work is supported by the National Natural Science Foundation of China $ ( $Grant Nos. 62173254 and 61773294$ ) $ , and by the National Key Research and Development Program of China $ ( $No.2020YFA0714200$ ) $.
\end{acknowledgements}

\section*{Compliance with ethical standards}

\textbf{Conflict of interest} The authors declare that they have no conflict of interest.


\begin{thebibliography}{}
	
	\bibitem{Nishi2020}
	Nishiura H, Jung S, Linton N M, et al. The extent of transmission of novel coronavirus in Wuhan, China, 2020[J]. Journal of clinical medicine, 2020, 9(2): 330.
	
	\bibitem{Hui2020}
	Hui D S, Azhar E I, Madani T A, et al. The continuing 2019-nCoV epidemic threat of novel coronaviruses to global health—The latest 2019 novel coronavirus outbreak in Wuhan, China[J]. International journal of infectious diseases, 2020, 91: 264-266.
	
	\bibitem{Tang2020}
	Tang B, Wang X, Li Q, et al. Estimation of the transmission risk of the 2019-nCoV and its implication for public health interventions[J]. Journal of clinical medicine, 2020, 9(2): 462.
	
	\bibitem{Liu2019}
	Liu J, Wu X, Lü J, et al. Infection-probability-dependent interlayer interaction propagation processes in multiplex networks[J]. IEEE Transactions on Systems, Man, and Cybernetics: Systems, 2019, 51(2): 1085-1096.
	
	\bibitem{Gross2006}
	Gross T, D’Lima C J D, Blasius B. Epidemic dynamics on an adaptive network[J]. Physical review letters, 2006, 96(20): 208701.
	
	\bibitem{Newman2002}
	Newman M E J. Spread of epidemic disease on networks[J]. Physical review E, 2002, 66(1): 016128.
	
	\bibitem{Zhu2020}
	Zhu S, Zhou J, Yu X, et al. Bounded synchronization of heterogeneous complex dynamical networks: A unified approach[J]. IEEE Transactions on Automatic Control, 2020, 66(4): 1756-1762.
	
	\bibitem{Wei2018}
	Wei X, Wu X, Chen S, et al. Cooperative epidemic spreading on a two-layered interconnected network[J]. SIAM Journal on Applied Dynamical Systems, 2018, 17(2): 1503-1520.
	
	\bibitem{Sun2018}
	Sun M, Small M, Lee S S, et al. An exploration and simulation of epidemic spread and its control in multiplex networks[J]. SIAM Journal on Applied Mathematics, 2018, 78(3): 1602-1631.
	
	\bibitem{Wang2019}
	Wang Z, Guo Q, Sun S, et al. The impact of awareness diffusion on SIR-like epidemics in multiplex networks[J]. Applied Mathematics and Computation, 2019, 349: 134-147.
	
	\bibitem{Granell2013}
	Granell C, Gómez S, Arenas A. Dynamical interplay between awareness and epidemic spreading in multiplex networks[J]. Physical review letters, 2013, 111(12): 128701.
	
	\bibitem{Bert2020}
	Bertozzi A L, Franco E, Mohler G, et al. The challenges of modeling and forecasting the spread of COVID-19[J]. Proceedings of the National Academy of Sciences, 2020, 117(29): 16732-16738.
	
	\bibitem{Ivorra2020}
	Ivorra B, Ferrández M R, Vela-Pérez M, et al. Mathematical modeling of the spread of the coronavirus disease 2019 (COVID-19) taking into account the undetected infections. The case of China[J]. Communications in nonlinear science and numerical simulation, 2020, 88: 105303.
	
	\bibitem{NiuR2020}
	Niu R, Wong E W M, Chan Y C, et al. Modeling the COVID-19 pandemic using an SEIHR model with human migration[J]. IEEE Access, 2020, 8: 195503-195514.
	
	\bibitem{Zhan2020}
	Zhan C, Tse C K, Fu Y, et al. Modeling and prediction of the 2019 coronavirus disease spreading in China incorporating human migration data[J]. Plos one, 2020, 15(10): e0241171.
	
	
	\bibitem{Liw2020}
	Li W, Zhou J, Lu J. The effect of behavior of wearing masks on epidemic dynamics[J]. Nonlinear Dynamics, 2020, 101(3): 1995-2001.
	
	\bibitem{Pastor2003}
	Pastor-Satorras R, Vespignani A. Epidemics and immunization in scale-free networks[J]. Handbook of Graphs and Networks, Wiley-VCH, Berlin, 2003.
	
	\bibitem{Zhou2011}
	Zhou X, Cui J. Analysis of stability and bifurcation for an SEIV epidemic model with vaccination and nonlinear incidence rate[J]. Nonlinear Dynamics, 2011, 63(4): 639-653.
	
	\bibitem{Lv2020}
	Lü W, Ke Q, Li K. Dynamical analysis and control strategies of an SIVS epidemic model with imperfect vaccination on scale-free networks[J]. Nonlinear Dynamics, 2020, 99(2): 1507-1523.
	
	\bibitem{Lwq2021}
	Li W, Zhou J, Jin Z, et al. The combination of targeted vaccination and ring vaccination[J]. Chaos: An Interdisciplinary Journal of Nonlinear Science, 2021, 31(6): 063108.
	
	
	
	
	
\end{thebibliography}
\end{document}